\pdfoutput=1
\documentclass[journal]{IEEEtran}
\usepackage{cite}

\ifCLASSINFOpdf
	\usepackage[pdftex]{graphicx}
	\graphicspath{{img/}}
	\DeclareGraphicsExtensions{.pdf,.jpeg,.png}
\else
	\usepackage[dvips]{graphicx}
	\graphicspath{{img/}}
	\DeclareGraphicsExtensions{.eps}
\fi

\usepackage{array}
\usepackage[cmex10]{amsmath}
\usepackage{mathtools}
\usepackage{mdwmath}
\usepackage{mdwtab}
\usepackage{color}
\usepackage[caption=false,font=footnotesize]{subfig}
\usepackage{stfloats}
\usepackage{eqparbox}
\usepackage{url}
\usepackage{amsfonts}
\usepackage{mathabx}
\usepackage{color}
\usepackage{etoolbox}
\usepackage[load=prefixed]{siunitx}
\usepackage{bbm}
\usepackage{courier}

\renewcommand{\d}{\mathrm{d}}

\newcommand{\intervo}[2]{\mathopen{]}#1\,;#2\mathclose{[}}

\newcommand{\xm}{x_{\mathrm{M}}}
\newcommand{\Cres}{C_{\mathrm{res}}}

\newcommand{\fz}{u_{\mathrm{0}}}

\newcommand{\xz}{x_{\mathrm{0}}}
\newcommand{\xcom}{x_{\mathrm{com}}}

\newcommand{\Cpm}{C_{\mathrm{\pm}}}
\newcommand{\Cm}{C_{\mathrm{-}}}
\newcommand{\Cp}{C_{\mathrm{+}}}

\newcommand{\Vres}{V_{\mathrm{res}}}

\newcommand{\Vp}{V_{\mathrm{+}}}
\newcommand{\Vm}{V_{\mathrm{-}}}

\newcommand{\xin}{\xi_{\mathrm{n}}}
\newcommand{\xieven}{\xi_{\mathrm{2 n}}}
\newcommand{\xiodd}{\xi_{\mathrm{2 n + 1}}}
\newcommand{\n}{\mathrm{n}}
\newcommand{\f}{\mathrm{f}}

\begin{document}
	\bstctlcite{IEEEexample:BSTcontrol}

	\title{\vspace{-0.2\baselineskip}Electrostatic Near-Limits Kinetic Energy\\Harvesting from Arbitrary Input Vibrations}%

	\author{Armine Karami, %
	J\'er\^ome Juillard, 
	Elena Blokhina, %
	Philippe Basset and %
	Dimitri Galayko
    
    \vspace{-0.5cm}
	
    \thanks{A. Karami is with the chair of econophysics and complex systems, LadHyX, Ecole Polytechnique, Palaiseau, France 
    \mbox{(e-mail: armine.karami@ladhyx.polytechnique.fr).}}
    \thanks{D. Galayko is with Sorbonne Universit\'e, LIP6, F-75005, Paris, France}%
	\thanks{J. Juillard is with GEEPS, CentraleSupélec, University of Paris-Sud, France.}
	\thanks{E. Blokhina is with the School of Electrical, Electronic and Communications Engineering, University College Dublin, Ireland.}
    \thanks{P. Basset is with Universit\'e Paris-Est, ESYCOM, ESIEE Paris, France.}}

	\markboth{}%
	{Karami \MakeLowercase{\textit{et al.}}: Electrostatic Near-Limits Kinetic Energy Harvesting from Arbitrary Input Vibrations}%

	\maketitle

	\begin{abstract}

		The full architecture of an electrostatic kinetic energy harvester (KEH) based on the concept of near-limits KEH is reported. This concept refers to the conversion of kinetic energy to electric energy, from environmental vibrations of arbitrary forms, and at rates that target the physical limits set by the device's size and the input excitation characteristics. This is achieved thanks to the synthesis of particular KEH's mass dynamics, that maximize the harvested energy. Synthesizing these dynamics requires little hypotheses on the exact form of the input vibrations. In the proposed architecture, these dynamics are implemented by an adequate mechanical control which is synthesized by the electrostatic transducer. An interface circuit is proposed to carry out the necessary energy transfers between the transducer and the system's energy tank. A computation and finite-state automaton unit controls the interface circuit, based on the external input and on the system's mechanical state. The operation of the reported near-limits KEH is illustrated in simulations which demonstrate proof of concept of the proposed architecture. A figure of $68\%$ of the absolute limit of the KEH's input energy for the considered excitation is attained. This can be further improved by complete system optimization that takes into account the application constraints, the control law, the mechanical design of the transducer, the electrical interface design, and the sensing and computation blocks.  
	\end{abstract}
	

	\vspace{-1\baselineskip}
	\section{Introduction}
	\label{sec:intro}

		\IEEEPARstart{K}{inetic} energy harvesting is increasingly investigated as a candidate technique to replace or increase the lifespan of batteries of miniaturized electronic systems. It consists in converting part of the mechanical energy of the system's surroundings vibrations to electrical energy, using a dedicated subsystem called a kinetic energy harvester (KEH). The size and mass of KEHs are small, so that the energy conversion does not impede with the system's surroundings, and yet the amount of harvested energy is enough to supply a low-power electronic system such as a wireless sensor. Typical applications include healthcare or structure monitoring. 

		A KEH is composed of a mechanical subsystem whose role is to capture kinetic energy from the vibrations that it is submitted to, and of an electromechanical subsystem whose role is to convert the captured mechanical energy into electrical energy and to supply a load. The mechanical part is at least composed of an inertial mass. The electromechanical part itself can be seen as a combination of a physical transduction device (e.g., a piezoelectric, electromagnetic or electrostatic transducer) associated with an electronic circuit that is responsible for the electrical conditioning of the transducer. This interface is also responsible for putting the energy in a form that can be used by the load (e.g., suitable DC voltage supply).

		The first works addressing inertial, one degree of freedom (1-dof) KEHs were focused on maximizing the amplitude of the motion of the mobile mass. Resonant mechanical systems have often been considered, and that restricted the applications of small-scale KEH to the cases where the source of the kinetic energy is narrowband vibrations of rather high frequencies (tens up to thousands hertz). The general effort has then been to improve the performances of KEHs, both in terms of converted energy and compatibility with low-frequency or large-bandwidth vibrations. To this end, the purposeful introduction of mechanical nonlinearities was studied. These include, for example, bistable KEHs \cite{cottone2009nonlinear}, or KEHs using frequency-up conversion \cite{lu2016batch}. The optimization was often done for vibration inputs of harmonic form, although some works report their study under white-Gaussian noise inputs \cite{halvorsen2008energy}. In these approaches, little attention was given to further optimization through the electrical interface. However, the electromechanical nature of a KEH system makes it inevitable to take it into account for full analysis and optimization. That is why for all types of transduction mechanisms, KEHs with smart electrical interfaces were studied \cite{lefeuvre2005piezoelectric,miller2016maximum,dicken2012power,yang2016reversible,karami2017series,zhao2017synergy,aghakhani2017equivalent}. These interfaces were designed to optimize the energy conversion, while at the same time providing the harvested energy to the load in a suitable form. The KEHs using these interfaces are often optimized for harmonic inputs, although some works study their operation for other types of inputs \cite{heit2016framework,blokhina2012limit,wu2015stochastic,badel2005efficiency,scruggs2012multi,scruggs2010causal}.


		In addition, theoretical works investigated the physical limits of kinetic energy conversion. Some of these works considered generic KEH architectures, yielding fundamental limits \cite{halvorsen2013architecture,heit2016framework}. Others works analyzed specific transduction mechanisms and interface electronics \cite{blokhina2012limit,mitcheson2012maximum} and yielded more conservative limits. In between, the important work \cite{mitcheson2004architectures} encompasses a large range of transduction mechanisms and conditioning electronics and proposed a categorization of a large part of the KEH architectures that were previously reported or reported since then.
		Most of these works studied the limits for specific types of input vibrations, e.g., harmonic \cite{halvorsen2013architecture,blokhina2012limit}, bi-harmonic, frequency swept \cite{heit2016framework}, or even white-Gaussian excitations for the specific case of KEH architectures using linear transduction mechanisms and resistive load \cite{halvorsen2008energy}. The obtained limits depend on the device size as well as the characteristic parameters of the input excitation.

          All these works noticed that the converted energy can be increased by appropriate choice of the dynamics of the mobile mass. Moreover, there exist an absolute upper limit of the power which can be harvested. This limit and the corresponding optimal dynamics, which we describe in sec. \ref{sec:principles}, are independent of the transduction mechanism and are only determined by the system size and the profile of the external vibrations.  The work \cite{mitcheson2004mems} was among the first studies that explicitly reported on an architecture attempting to implement such dynamics, but the constrained form of the control was not suited to maximize the converted energy from inputs of irregular form.
		Later in \cite{Miao2006}, the same group mentioned that a similar system that would dynamically adapt to the input excitation is needed in order to approach the limits of energy conversion from more general types of input excitations. Such a system was sketched more recently in \cite{hosseinloo2015non}. It uses a dedicated control force created by magnetic transduction to implement the needed dynamics for a piezoelectric KEH. This latter work also shows how such controlled KEHs outperform resonant-based and bistable nonlinearity-based KEHs in the case of irregular input vibrations. Yet, the quantitative description of the control and of the needed electrical interface was not reported in this study and is yet to be addressed. 

          In this paper, we address a quantitative description of a system implementing the optimal dynamic for the mobile mass, with the aim to approach the physical limit of the conversion energy.   
		We will refer to such a KEH as a "near-limits KEH". We propose an its implementation based on electrostatic transduction. Although other type of transduction may also be suitable for such a system, we beleive that electrostatic transduction is better compatible with low scale devices (through the MEMS technologies) and offers more precise control of the mechanical forces. The optimal dynamics are synthesized directly by the transducer force used for the electromechanical energy conversion, using an appropriate control force associated with a suited interface circuit. 
          
          The plan of the paper is as follows. First, in Sec. \ref{sec:principles}, we outline the principles of near-limits KEH. Then, in Sec. \ref{sec:electrostatic_implementation}, we detail the electrostatic near-limits KEH architecture. In Sec. \ref{sec:sizing_simulations}, we report the results of the electromechanical simulation of the proposed architecture. The results illustrate its operation submitted to realistic vibrations recorded on a running human.
          The proposed architecture is validated by a simulation of a hybrid behavioural/Spice model.

	\vspace{-0.5\baselineskip}
	\section{Principles of near-limits KEH}
	\label{sec:principles}

		This section is devoted to summarize the theoretical foundations of the near-limits kinetic energy harvesting. The sec. \ref{subsec:principles_non_net} derives the mobile mass dynamics maximizing the converted energy in the context of an ideal lossless KEH model. This derivation is essentially the same as what is reported in \cite{hosseinloo2015fundamental}: we report it again for the sake of completeness.
          In sec. \ref{subsec:general_optimization_problem} we complete the near-limit KEH model by discussing on the maximization of the harvested energy in the presence of losses. 

		\vspace{-0.75\baselineskip}
		\subsection{Statement of the principles for an idealized KEH model}
		\label{subsec:principles_non_net}

			\begin{figure}[htbp]
				\centering
				\includegraphics[width=.49\textwidth]{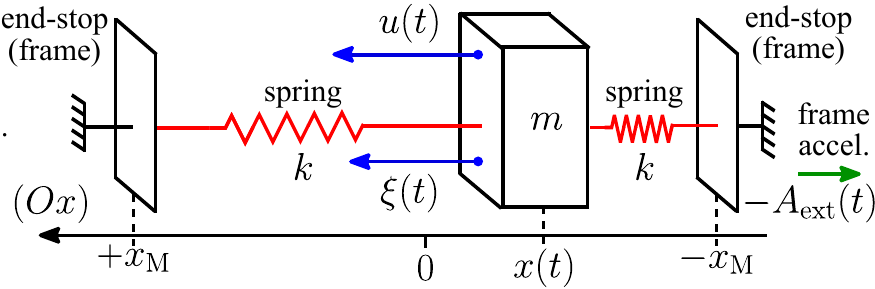}
				\caption{A generic model for a 1-dof kinetic energy harvester.}
				\label{fig:systemgeneric}
			\end{figure}

               A generic, lineic, inertial kinetic energy harvester (KEH) system is depicted in Fig. \ref{fig:systemgeneric}. The KEH is composed of a mass which is supposed to be free to move along the 0x axis attached to the vibrating frame (the vibrating box), and $x(t)$ denotes its position along this axis at time $t$, as depicted in the figure. The mass is attached to the frame's end-stops using a suspension modeled as a linear spring of stiffness $k$. The spring is added in the generic ``mass in a box'' model to account for the suspension of the mass in the 1-dof system. The box in which the mass lies has limited dimensions, and as a result, the mass displacement inside the frame is constrained between positions \mbox{$\pm \xm$}. The box has an acceleration of \mbox{$-A_{\mathrm{ext}}(t)$}: hence, when the 2nd Newtonian law is written in the Ox frame, the mass is submitted to an input force \mbox{$\xi(t) = m A_{\mathrm{ext}}(t)$}.

			The mass is also submitted to a force $u(t)$. In this generic model, all that is assumed is that this force is of electrical origin. It is hence responsible for the mechanical to electrical energy conversion. The details of its implementation depend on the electrical to mechanical transduction mechanism that is used to generate it (e.g., electromagnetic, piezoelectric, electrostatic), as well as on the actual mechanical and electrical subsystems of the KEH (see Sec. \ref{subsec:general_optimization_problem}).

			The ODE describing the dynamics of this system is:
			\begin{equation}
				m \ddot{x}(t) = \xi(t) + u(t) - k x(t).
				\label{eq:ode_model}
			\end{equation}
			where the overdot denotes differentiation with respect to time. The system's energy balance reads:
			\begin{equation}
				\begin{aligned}
					\int_{t_0}^t &\xi(s) \dot{x}(s) \d s = - \int_{t_0}^t u(s) \dot{x}(s) \d s \\&+ 1/2\cdot(m(\dot{x}^2(t) - \dot{x}^2(t_0)) + k (x^2(t) - x^2(t_0))).
				\end{aligned}
				\label{eq:energy_fluxes}
			\end{equation}
			The LHS is the \emph{input energy} in the system, and the first term of the RHS is the \emph{converted energy} from the mechanical to the electrical domain by the force $u(t)$, or \emph{converted energy}.
			Let us also suppose that \mbox{$|x(t)| \leq \xm$}. In this context, choosing a force $u(t)$ that maximizes the converted energy:
			\begin{equation}
				\begin{aligned}
					\begin{cases}
						\max\limits_u&\left(-\int_{t_0}^t u(s) \dot{x}(s) \d s\right)\\
						\text{s.t. }&m \ddot{x}(t) = \xi(t) + u(t) - k x(t)\\
						&|x(t)| \leq \xm
					\end{cases}
				\end{aligned}
			\end{equation}
			is done by maximizing the input energy  $\int_{t_0}^t \xi(s)\dot{x}(s)\d s$, as (\ref{eq:energy_fluxes}) suggests.
			This, in turn, is done by finding the mass trajectory $x_{\xi}(t)$ that maximizes this integral for any input excitation $\xi(t)$.

			Let us first seek the trajectory $x_{\xi}(t)$ maximizing the input energy within the set $X$ of piecewise-differentiable, bounded functions. Let us rewrite
			\begin{equation}
				\int_{t_0}^t \xi(s)\dot{x}(s)\d s = - \int_{t_0}^t \dot{\xi}(s) x(s) \d s + \xi(t)x(t) - \xi(t_0)x(t_0)
			\end{equation}
			The input excitation and the mass position in the box being bounded, the two last terms of the RHS are bounded. Hence, our goal being to optimize the mean converted power, we just need to maximize the integral on the RHS.

			In the rest of the paper, the consecutive local extrema of $\xi(t)$ are represented by the sequence $(\xin)$. The corresponding times at which they happen are represented by the sequence $(t_n)$. The maximums are of even indexes $(\xieven)$, while the minimum have odd indexes $(\xiodd)$. As \mbox{ $-\int_{t_0}^t \dot{\xi}(s)x(s)\d s \leq \xm \int_{t_0}^t |\dot{\xi}(t)|\d t$}, it comes that:
			\begin{align}
					&\max_{x \in X} \left(\int_{t_0}^t {\xi}(s)\dot{x}(s) \d s\right)\!=\!2 \xm\left(\,\,\label{eq:max_energy}\sum_{\mathclap{\,\,\,\,\{t_{2n} \leq t\}}}\,\xieven -  \sum_{\mathclap{\,\,\{t_{2n+1} \leq t\}}}\,\xiodd\right),\\
					&\text{ attained at: }x_{\xi}(t)=\begin{cases}
						 \xm, t \in \intervo{t_{2 n}}{t_{2 n + 1}},\\
					   - \xm, t \in \intervo{t_{2 n + 1}}{t_{2 (n + 1)}}
					\end{cases}
					\notag
			\end{align}

\begin{figure}[htbp]
				\centering
				\includegraphics[width=.49\textwidth]{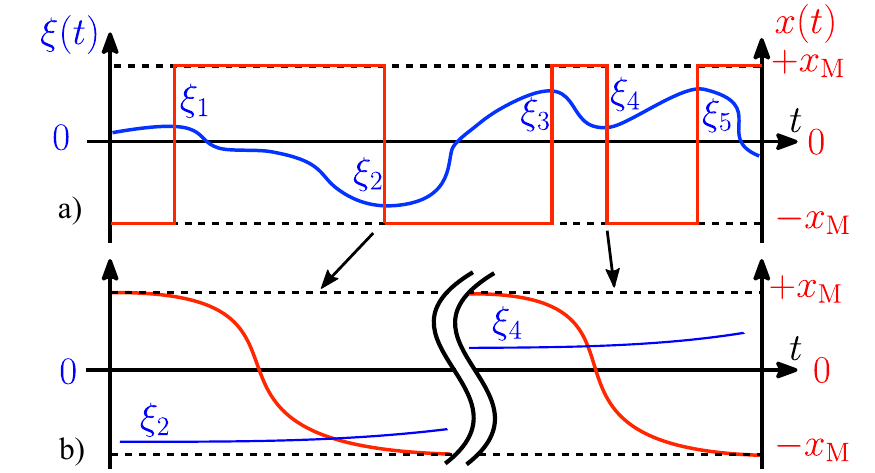}
				\caption{Example of a smooth energy-maximizing trajectory $x_\xi(t)$ for an example of input $\xi(t)$. Two position switching events are time-magnified.}
				\label{fig:optimal_trajectory}
			\end{figure}

			The motion $x_{\xi}(t)$ lacks regularity, but a smooth trajectory pointwisely approaching it will result in the input/converted energy approaching the maximum. Such a trajectory has to switch from one end-stop $\pm \xm$ to the other end-stop $\mp \xm$ as soon as a maximum/minimum of the input force is detected. This switching has to be done as fast as possible, to happen while the external force is constant and equal to the detected maximum/minimum value. In addition, the mobile mass must arrive at the other end-stop with ideally zero velocity. If the mass arrives at the other end with a non-zero velocity, the outcome will depend on the physical implementation of the end-stop (the model (\ref{eq:ode_model}) supposes constrained motion with no definition of the dynamics out of the displacement limits). Bouncing against the end-stop can make the trajectory deviate from the trajectory $x_\xi(t)$. Also, if the collisions with the end-stops are inelastic, some energy will be wasted. Finally, in between a maximum and a minimum (resp. a minimum and a maximum) of $\xi(t)$, the mobile mass must be kept still at $-\xm$ (resp $\xm$). An example summarizing these requirements is depicted in Fig. \ref{fig:optimal_trajectory}. From now on, $x_\xi(t)$ will denote this smoothened target trajectory.

			In the setting of this simple model, maximizing the converted energy from a given input excitation is done by implementing the control force $u(t)$ that results in $x_\xi(t)$. 
			The description of an electrostatic KEH architecture implementing this control, while recovering the converted energy, is the objective of this paper. Before proceeding to the details of the architecture in Sec. \ref{sec:electrostatic_implementation}, the following discussion highlights the differences between the strategy just presented and the formal problem of harvested energy maximization.

		\vspace{-0.75\baselineskip}
		\subsection{On the difference between near-limits KEH and the formal problem of maximization of the harvested energy}
		\label{subsec:general_optimization_problem}

			In the hypothesis of (\ref{eq:ode_model}), the input and converted energy are equal. Thus, the dynamics maximizing the input energy also maximize the converted energy. Yet, a realistic model has to include mechanical friction phenomena, e.g., air damping or inelastic collisions with the end-stops, due to possible inaccuracy in the control implementation. This makes the converted energy smaller than the input energy.

			Moreover, the ultimate goal of KEH is not the mere maximization of the converted energy, but rather of the \emph{harvested energy}. We define this quantity as the part of the converted energy that ends up in the system's electrical energy tank, in a form usable by an electrical load. The harvested energy is smaller than the converted energy, because of the dissipation phenomena that occur in the electrical domain. These losses are the source of the control generation cost. Indeed, to implement $u(t)$, the system must invest energy from its electrical energy tank into the transducer.
			This energy transfer is done using an interface circuit. This circuit has sources of energy dissipation that result in the loss of part of the invested energy during the transfer. The same phenomenon occurs when the interface circuit transfers the invested energy from the transducer back to the energy tank with the overhead of converted energy. Also, the interfacing circuit can be driven by a control unit whose decisions are computed as a function of the desired $u(t)$, as well as, at least, the input $\xi(t)$. The energy consumed by this computing and sensing is a part of the dissipation of the system. 

			Given these considerations, the trajectory $x_\xi(t)$ only results in the maximum of harvested energy in the case of an idealized system. If the energy dissipation sources are taken into account, then not only the harvested energy is decreased by the amount of dissipated energy, but also $x_\xi(t)$ does not correspond to the trajectory maximizing the harvested energy.

			For the sake of completeness and to highlight the differences with the strategy derived in Sec. \ref{subsec:principles_non_net}, let us give a general formulation of the problem of harvested energy maximization. A KEH system implements the force $u(t)$ through a transduction mechanism that depends on the mechanical and electrical states of the KEH. The electrical state refers to the currents $\mathbf{I}$, and voltages $\mathbf{V}$, of the electrical circuit connected to the transducer. This can be summarized by a relation \mbox{$u(t) = \mathbf{\Phi}(\mathbf{I},\mathbf{V},\dot{x},x)$}, where $\mathbf{\Phi}$ describes a particular transduction mechanism. A force $u(t)$ is then implemented by adequately controlling the currents and voltages of the interface circuit to achieve the energy transfers between the electrical energy tank and the transducer. This interface circuit can be described by an implicit relation $\mathbf{\Psi}(\dot{\mathbf{I}}, \dot{\mathbf{V}}, \mathbf{V},\mathbf{I},\dot{x},x,t) = \mathbf{0}$. Here, $\mathbf{\Psi}$ consists in a set of differential-algebraic relations that describe the circuit's evolution, constrained by the electrical network and components laws. The time dependence accounts for the control of the interface resulting in $u(t)$. The mechanical suspension can be represented by a force $-k(x)x$, possibly including nonlinear stiffness effects. Dissipative phenomena in the mechanical domain can be represented by a force $-c(\dot{x},x)\dot{x}$. Both $k(x)$ and $c(\dot{x},x)$ are likely to be piecewise-defined to model both the region where the mass is within $\pm \xm$ and the collision between the mass and the end-stops \cite{truong2015verified}. Note that some works reported power-extracting end-stops \cite{phule2013powerextracting}, which could possibly relax the requirements on the control $u(t)$. The control generation cost $C_e(\mathbf{\Psi})$ arises from the aforementioned dissipation phenomenon in the electrical domain. It depends on the chosen control function and on the electrical interface used to implement it, all of which are described by $\mathbf{\Psi}$.

			Given the previous notations and an input $\xi(t)$, the generic problem of harvested energy maximization is the selection, through the electrical interface described by $\mathbf{\Psi}$, of the adequate control. The formal problem reads:
			\begin{equation}(P)
				\begin{cases}
					\max\limits_{\substack{\mathbf{\Psi}}}& \{-\int_{t_0}^t u(s)\dot{x}(s)\d s - C_e(\mathbf{\Psi}) - \int_{t_0}^t c(\dot{x},x)\dot{x}\d s\}\\
					&
					m \ddot{x}(t) = \xi(t) + u(t)\notag - k(x) x(t) - c(\dot{x},x)\\
					&u = \Phi(\mathbf{I}, \mathbf{V}, \dot{x}, x),\quad\mathbf{\Psi}(\dot{\mathbf{I}}, \dot{\mathbf{V}}, \mathbf{I}, \mathbf{V}, \dot{x}, x, t) = 0
				\end{cases}.
			\end{equation}

			Solving the problem $(P)$, given a predetermined transduction mechanism, consists in choosing the optimal electrical subsystem $\mathbf{\Psi}$ and ``simultaneously'', through it, the optimal control policy $u(t)$. Different special cases of the problem $(P)$ for various types of inputs were recently reported in the literature. For example, in \cite{halvorsen2016optimal}, the problem is solved in the setting of a determined form of input excitation, with a constrained electrical interface (resistive load) and a linear transduction mechanism. The control is done through the load value and the mechanical stiffness. The works in \cite{hosseinloo2015fundamental,haji2015optimal} also report on different optimization and optimal control problems that are particular cases of the problem $(P)$.

			Yet, designing a KEH maximizing the harvested energy -- in the sense that it implements a solution to the problem $(P)$ -- requires, in general, that $\xi(t)$ is known is advance. A simple example of this is reported in \cite{halvorsen2013architecture} where a linear mechanical friction force is taken into account. In this case, the shortfall is only in the converted energy relatively to the input energy, as the electrical interface that could further decrease the harvested energy is not considered. Even in this simple case, the paper shows that the dynamics maximizing the converted energy can only be implement if the form of the external input is known (a harmonic excitation in the case of the cited study).

			In contrast, the present work is concerned with implementing an architecture achieving the control strategy described in \ref{subsec:principles_non_net}. This can be done without \emph{a priori} knowledge on the exact form of $\xi(t)$. This is because it relies on the synthesis of the motion $x_\xi(t)$, which merely requires a real-time detection of the extrema of $\xi(t)$. This detection can be done by a causal system, provided that (i) the detection is fast enough so that the value of $\xi(t)$ is still close enough to it (ii) the resolution of the sensor used to measure $\xi(t)$ is high enough. In practice, the next sections will show that some limited information about the input is necessary to correctly synthesize $x_\xi(t)$.

			Although our strategy does not yield the upper limit of harvested energy, we can suppose as a working hypothesis that this limit can be approached by the strategy of implementing $x_\xi(t)$ as the costs \mbox{$C_\mathrm{e} + \int_{t_0}^t c(\dot{x},x)\dot{x}\d s$} approach zero. This reasonable hypothesis is supported by \cite{halvorsen2013architecture}, where the trajectory maximizing the converted energy becomes closer to $x_\xi(t)$ (for a harmonic $\xi(t)$) as the damping parameter becomes sufficiently small. Thus if we manage to make the control generation cost to implement $x_\xi(t)$ small, the harvested energy will approach the upper limit yielded by the problem $(P)$.

	\vspace{-0.5\baselineskip}
	\section{Architecture of an electrostatic near-limits kinetic energy harvester}
	\label{sec:electrostatic_implementation}

		\begin{figure}[htbp]
			\centering
			\includegraphics[width=.49\textwidth]{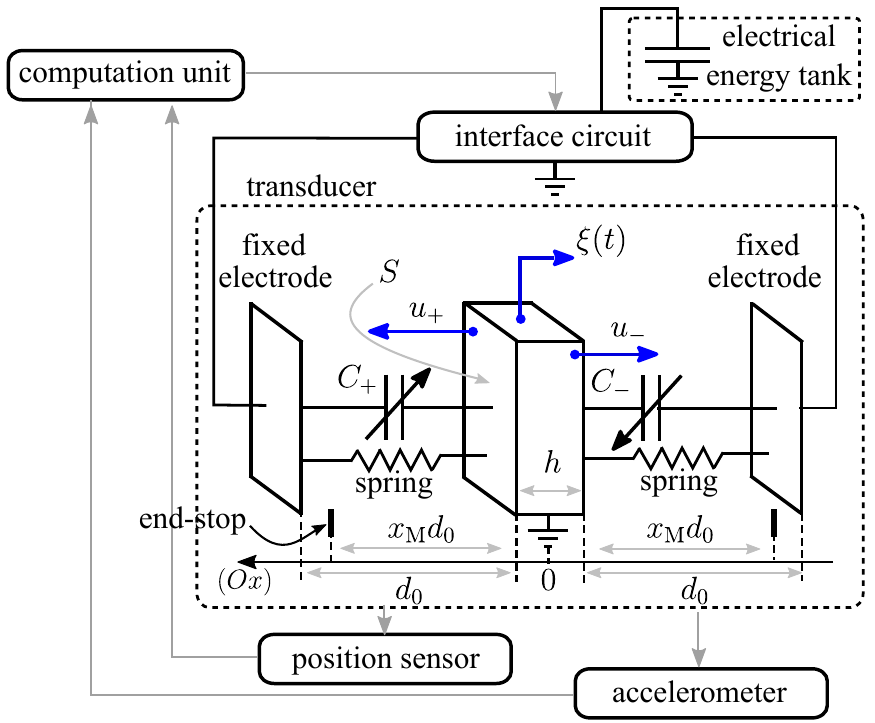}
			\caption{Overview of the architecture for a near-limits electrostatic KEH.}
			\label{fig:electrostatic_nearlimits_keh}
		\end{figure}
		
		In this section, the architecture of a near-limits KEH based on electrostatic transduction is described. Specifically, the KEH implements a control synthesizing the trajectory $x_\xi(t)$ with limited hypotheses on the input excitation.
		The architecture comprises an electrical interface that is suited to carry out the necessary energy exchanges between the KEH's electrical energy tank and the electrostatic transducer to synthesize the control whilst recovering the converted energy. The electrical interface is driven by a sensing and computation unit which revolves around a finite-state automaton.
		A block-diagram of the proposed architecture is depicted in Fig. \ref{fig:electrostatic_nearlimits_keh}.

		\vspace{-0.75\baselineskip}
		\subsection{Electrostatic transducer}
		\label{subsec:electrostatic_transduction}


			An electrostatic electromechanical transducer is a variable capacitor. In our case, we are interested in transducers whose change of geometry make their capacitance vary. This change of geometry happens thanks to the movement of one of the mobile electrodes of the transducer. This movement is constrained in amplitude between the end-stops, and on one axis, as in the generic model of Sec. \ref{subsec:principles_non_net}.

			In this context, the transduction is materialized by a mechanical force between the transducer's electrodes that depends on the voltage across them. It is supposed that one of the electrodes is fixed to the frame, the other being attached to the mobile mass of the KEH. The transducer force acting on the mobile mass has the expression:
			\begin{equation}
				u(t) = \frac 1 2 \cdot V^2(t) \cdot \frac{\partial C}{\partial x}(x(t))
			\end{equation}
			The term $\partial C/\partial x$ depends on the transducer's geometry, and $V(t)$ is the voltage across the transducer.

			To generate forces of both signs on the axis $(O x)$, a differential transducer composed of two electrostatic transducers $C_{+}$ and $C_{-}$ must be used. Each of these transducers has one of its electrodes attached to the frame's ends, and the other electrode attached to the KEH's mass. A schematic view of this geometry is depicted in Fig. \ref{fig:electrostatic_nearlimits_keh}. The force $u$ is then the superposition of the two forces:
			\begin{equation}
				u(t) = u_+(t) - u_-(t)
			\end{equation}
			with
			\begin{equation}
				\begin{aligned}
					u_{\mathrm +}(t) &= \frac 1 2 \cdot {\Vp}^2(t) \cdot \frac{\partial \Cp}{\partial x}(x(t))\\
					u_{\mathrm -}(t) &= \frac 1 2 \cdot {\Vm}^2(t) \cdot \frac{\partial \Cm}{\partial x}(x(t))\\
				\end{aligned}
			\end{equation}

			The geometry that we choose for our near-limits KEH is a gap-closing geometry. This means that the capacitance variation of the transducers results from their mobile electrode sliding along the axis that is normal to their plane (also called gap-closing or out-of-plane geometry). 
			
			Let us define \mbox{$\kappa = \epsilon_0 S / {d_0}^2$}, a constant that depends on the device's geometry: the gap $d_0$ at \mbox{$x = 0$}, the total surface of the capacitance (transducer overlapping area) $S$, and the dielectric constant $\epsilon_0$. Also, from now on, let us substitute all positions with their value normalized by dividing by $d_0$, i.e., substitute \mbox{$x(t)/d_0 \rightarrow x(t)$}, \mbox{$\xm/d_0 \rightarrow \xm$}. The capacitance of a gap-closing transducer is hence expressed as:
			\begin{equation}
				\begin{aligned}
					C_\pm(x(t)) = \frac{\kappa d_0}{(1 \mp x(t))},
				\end{aligned}
			\end{equation}
			and hence:
			\begin{equation}
				\begin{aligned}
					u_\pm(t) &= \frac 1 2\cdot {V_\pm}^2(t) \cdot \frac{\kappa}{  (1 \mp x(t))^2},
				\end{aligned}
				\label{eq:electrostatic_force}
			\end{equation}
			From (\ref{eq:electrostatic_force}), it comes that with gap-closing electrostatic transducers, when the charge of $\Cpm$ is kept constant through the displacement of the mass, the force that the corresponding transducer generates on it is constant. Specifically, if at position \mbox{$x(t_0) = x_0$}, the transducer is charged with \mbox{$Q_{\pm}(t_0) = V_{\pm}(t_0) \Cpm(x_0)$}, then the transducer force is constant across the displacement, as long as $Q_{\pm}(t)$ remains constant
			\begin{equation}
				\begin{aligned}
					\forall t, t_0 \leq t \leq t_1 \implies u_\pm(t) = u_{\pm}(t_0) = \frac{\kappa}{2} \frac{{V_{\pm}}^2(t_0)}{(1 \mp x_0)^2}\\
					t_1 = \mathrm{inf}\,\{\,t\geq t_0\,|\,Q_\pm(t) \neq Q_{\pm}(t_0)\,\}
				\end{aligned}
				\label{eq:force_vs_voltage}
			\end{equation}
			This is referred to as the ``constant-charge operation''. The corresponding energy in $\Cpm(x_0)$ when it is charged reads:
			\begin{equation}
				W_{0} = u_\pm(t_0) d_0 (1 \mp x_0)
				\label{eq:invested}
			\end{equation}

		\vspace{-0.75\baselineskip}
		\subsection{Control strategy}
		\label{subsec:ctrl_strategy}

			Let us now describe the control force $u(t)$ that synthesizes the trajectory $x_\xi(t)$ described in Sec. \ref{subsec:principles_non_net}. 


			\begin{figure}[htbp]
				\centering
				\includegraphics[width=.49\textwidth]{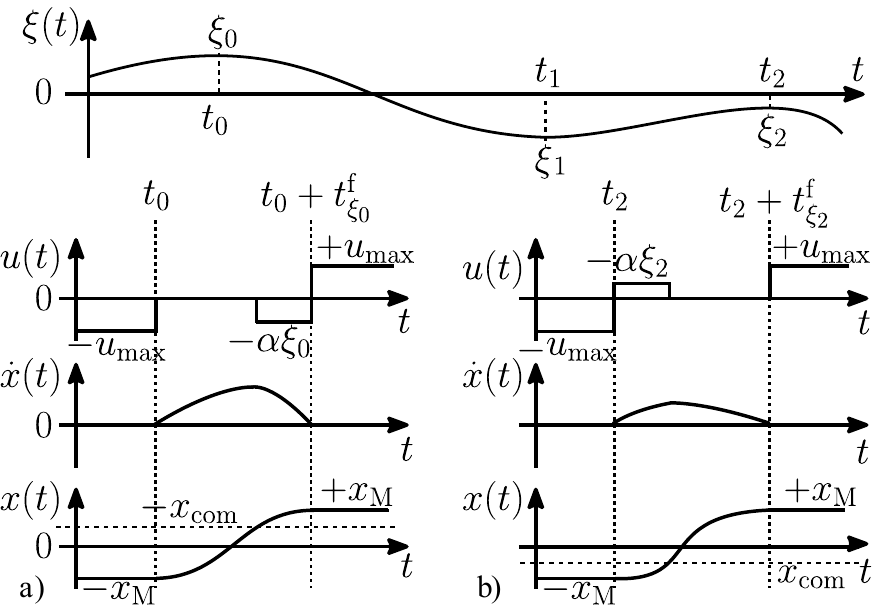}
				\caption{Detailed waveforms illustrating the proposed control strategy, on two examples of external force local maximum.}
				\label{fig:control_electrostatic}
			\end{figure}

			\subsubsection{Description of the control}
			\label{subsubsec:detailed_description_ctrl}

               The chosen control scheme can be explained as follows. Suppose that initially, the mass is located at $-\xm$ and $\xi(t)$ is increasing. The control $u(t)$ sustains a force value of $-u_\mathrm{max}$ so that the mass, on which $\xi(t)$ and \mbox{$-k x(t)$} act, is kept at $- \xm$ (thanks to the reaction force of the end-stop). Then, as soon as a maximum of $\xi(t)$ is sensed, two cases are possible: either the value of the maximum of $\xi(t)$ is positive, either it is negative. The first case is depicted in Fig. \ref{fig:control_electrostatic}.a. In this case, the force $u(t)$ is removed (set to zero), and the mass travels towards $+\xm$ under the action of $\xi(t)$, that is supposed varying slowly enough so that its value is equal to the value of the local maximum $\xi_0$. When the mass reaches a given position \mbox{$-\xcom \in \intervo{-\xm}{\xm}$} (normalized by division by $d_0$), the control force $u(t)$ takes a constant value, working negatively on the mass. This value is computed so that the mass arrives at \mbox{$+\xm$} with zero velocity.

               The other case is depicted in Fig. \ref{fig:control_electrostatic}.b, where the value $\xi_2$ of the maximum of $\xi(t)$ is negative. In this case, the force $u(t)$ first takes a constant value, and works positively on the mass so that it travels towards $+\xm$. When the mass reaches $\xcom$, the force $u(t)$ is removed. The value the positively working force is chosen so that the mass reaches \mbox{$+\xm$} with zero velocity. In both cases, as soon as the mass reaches \mbox{$-\xm$}, the control force takes the value $u_\mathrm{max}$ to keep the mass at $+\xm$. Then, when a minimum of $\xi(t)$ is sensed, the symmetrical sequence of actions happens, where we substitute \mbox{$\xcom \rightarrow -\xcom$}. 
				Formally, the control reads:
				\begin{equation}
					u(t) =
					\begin{cases}
						\tilde{u}_n(t), &\text{ if } t \in \intervo{t_\n}{t_\mathrm{n+1}+t^\f_{\xi_\mathrm{n}}}\\
						u_{\mathrm{max}}, &\text{ if } t \in \intervo{t_{2\n} + t^\f_{\xi_{2\n}}}{t_{2\n+1}}\\
						-u_{\mathrm{max}}, &\text{ if } t \in \intervo{t_{2\n+1}+t^\f_{\xi_{2\n+1}}}{t_{2(\n+1)}}
						\label{eq:controlscheme}
					\end{cases}
				\end{equation}
				where $\tilde{u}_n(t)$ is a solution to the position switching problem:%
				\begin{equation}(Q)
					\begin{cases}
						m d_0 \ddot{x}(t) = \xin + u(t)\notag - k d_0 x(t)\\
						(x(0), \dot{x}(0)) = (-x(t^\mathrm{f}_\n), \dot{x}(t^\mathrm{f}_\n))  = (\pm \xm, 0)\\
						|x(t)| \leq \xm
					\end{cases},
				\end{equation}
				and reads:
				\begin{equation}
					\begin{aligned}
						&\tilde{u}_n(t) = 
						\begin{cases}
								\begin{cases}
									\,\,\,\, 0 & x(t) < -\xcom \\
									\,\,\,\, -\alpha \xin & x(t) \geq -\xcom
								\end{cases}\text{ if } \xi_\n > 0,\\
								\begin{cases}
									\,\,\,\, -\alpha \xin & x(t) < \,\,\,\,\xcom\\
									\,\,\,\, 0 & x(t) \geq \,\,\,\,\xcom
								\end{cases}\text{ if } \xi_\n < 0
						\end{cases}\\
					&\text{where } \alpha \coloneqq  2\xm/(\xm + \xcom).
					\end{aligned}
				\end{equation}
				Here, $t^\mathrm{f}_\xi$ is the duration of the position switching from \mbox{$\pm \xm$} to \mbox{$\mp \xm$} for an extremum of value $\xi$. It reads:
				\begin{equation}
					\begin{aligned}
						&t^\mathrm{f}_\xi = \sqrt{\frac{m}{k}}\left(\delta + 
						\mathrm{acos}\left(
						\frac{|\xi| + k d_0 \xcom}{|\xi| + k d_0 \xm}
						\right)\!\right.+\\&\left.
						\!\!+ \mathrm{atan}\!\left(
						\frac{(\xm\!\!+\!\!\xcom)\sqrt{k d_0(\xm\!\!-\!\!\xcom)(2|\xi|\!\!+\!\!k d_0 (\xm\!\!+\!\!\xcom))}}{|\xi|(\xm\!\!-\!\!\xcom)\!\!-\!\!k d_0 \xcom(\xm\!\!+\!\!\xcom)}
						\right)\!\!\right),
					\end{aligned}
					\label{eq:duration_switching}
				\end{equation}
				where
				\begin{equation}
					\delta \!\coloneqq\!\!\begin{cases} \!\pi \text{ if}-\!\sqrt{1 \!\!+ \!\!\frac{4 k d_0 \xm |\xi|}{(k d_0 \xm + |\xi|)^2}} \!\!<\!\! 1 \!\!+\!\! \frac{2 k d_0 \xcom}{k d_0 \xm + |\xi|} \!\!<\!\! \sqrt{1 \!\!+\!\! \frac{4 k d_0 \xm |\xi|}{(k d_0 \xm + |\xi|)^2}} \\\!0 \text{ otherwise }\end{cases}
				\end{equation}
				and where \mbox{$u_{\mathrm{max}} > \xi_\mathrm{max} + k d_0 \xm$} (\mbox{$\xi_\mathrm{max} \coloneqq \max_t |\xi(t)|$}). The value of $t^\mathrm{f}_{\xi_\mathrm{n}}$ is needed to assess if the position switching is fast enough such that it can be considered that \mbox{$\xi(t) \approx \xi_{\mathrm{n}}$}. Otherwise, the control will lack accuracy and the implemented trajectory will deviate from $x_\xi(t)$, resulting in a shortfall in the converted energy. Estimating  $t^\mathrm{f}_{\xi_\mathrm{n}}$ requires some information about the input to which the system is going to be submitted. After noticing that $t^\mathrm{f}_\xi$ is a decreasing function of $\xi$, we may conservatively require that
				\begin{equation}
					\lim\limits_{\xi \to 0} t_\xi^\mathrm{f} < K/f_{\mathrm{max}},
					\label{eq:hyp1}
				\end{equation}
				for large enough $K$ depending on the tolerated error, and where $f_{\mathrm{max}}$ denotes the highest frequency component of $\xi(t)$ that has significant amplitude. Studies of specific application contexts of KEH can provide such information. Also, note that $t^\mathrm{f}_\xi$ decreases when $k$ is increased.

			\subsubsection{Energy considerations}
			\label{subsubsec:energy_considerations_ctrl}

				The conversion between mechanical and electrical energy occurs during the position switchings at the extrema of $\xi(t)$.
				In the case of a maximum of positive value or of a minimum of negative value (in these two cases, \mbox{$x(t_\n) \xi_\n < 0$}), the position switching results in the conversion of energy from the mechanical to the electrical domain. At each such extremum, from (\ref{eq:invested_diffsigns}), it comes that the energy to invest to implement $\tilde{u}(t)$ reads:
				\begin{equation}
					{W_{\mathrm{i,n}}}^- = 2 d_0 \xm\frac{1 - \xcom}{\xm + \xcom}\cdot|\xin|.
					\label{eq:invested_diffsigns}
				\end{equation}
				In the case of a maximum of negative value or of a minimum of positive value (i.e., \mbox{$x(t_\n) \xi_\n > 0$}), the position switching results in the conversion of energy from the electrical to the mechanical domain, and the next extremum $\xi_{\mathrm{n}+1}$ will necessarily verify \mbox{$\xi_{\mathrm{n}+1} x(t_{\mathrm{n}+1}) < 0$}. From (\ref{eq:invested_samesigns}), the invested energy reads:
				\begin{equation}
					{W_{\mathrm{i,n}}}^+ = 2 d_0 \xm\frac{1 + \xm}{\xm + \xcom}\cdot|\xin|.
				 	\label{eq:invested_samesigns}
				\end{equation}
				The converted energy between the mechanical and the electrical domain (denoted ${W_{\mathrm{c,n}}}^-$ for the case \mbox{$\xin x(t_\mathrm{n}) < 0$} and ${W_{\mathrm{c,n}}}^+$ for the case \mbox{$\xin x(t_\mathrm{n}) > 0$}) reads:
				\begin{equation}
					{W_{\mathrm{c,n}}}^\pm = \mp 2 d_0 \xm |\xi_{\mathrm{n}}|.
					\label{eq:recovered}
				\end{equation}
				A negative amount of converted energy means that the energy is converted from the electrical to the mechanical domain. Summing (\ref{eq:recovered}) over $n$, one obtains the expression (\ref{eq:max_energy}).

				For a practical implementation, an important parameter is the maximum voltage across the transducer. It is reached when the energy on the transducer $\Cpm$ is equal to the sum of invested and converted energy ((\ref{eq:invested_diffsigns}) and (\ref{eq:recovered})) for $\xi_\mathrm{max}$ at position \mbox{$\mp \xm$}. It reads
				\begin{equation}
					V_{\mathrm{max}}\!= 2\cdot(1+\xm)\sqrt{\frac {1}{\kappa}\frac{x_\mathrm{M}}{\xm+\xcom}\xi_\mathrm{max}}
					\label{eq:vmax}
				\end{equation}

			\subsubsection{Comments}
			\label{subsubsec:ctrl_comments}

				In Sec. \ref{subsec:general_optimization_problem}, it was highlighted that the optimal control in terms of harvested energy has to be implemented by simultaneously determining the mathematical control law and the interface that has to physically implement it. The problem $(Q)$ can itself be enriched with such model details and cost functionals, e.g., adding end-stops model or mechanical friction phenomenon and/or minimizing the energy dissipated in a determined interface circuit. This allows one to take into account the coupling between the choice of both the mathematical control law and the interface used to implement it, that $(P)$ already highlighted.

				The choice of the simple open-loop, piecewise-constant control (\ref{eq:controlscheme}) results from considering this entanglement on a less formal level. 
				A piecewise-constant control is easily implemented thanks to property of the transducer to generate a constant force at a constant charge (see Sec. \ref{subsec:electrostatic_transduction}). To generate such a control, it is enough to setup the transducer's charge at discrete time instants, and to keep the charge constant during the remaining time. This is easily achieved with switching interfaces described in the literature, and architecture we propose in this paper is based on these types of interfaces.

				In addition, the chosen piecewise-constant force minimizes the number of force updates (one update per position switching, which is the minimum number of commutations between constant values of the force that are necessary to solve $(Q)$). This is because each update will require the interface to perform an action with an associated energy cost. Moreover, one of the constant force values is $0$ in order to avoid additional energy investment, at the cost of a slower position switching. Increasing the invested energy to implement the position switchings results in greater losses in the interfacing circuit for the same value of converted energy per extremum (\ref{eq:recovered}), hence decreasing the harvested energy. To further decrease the invested energy and hence the cost of control generation, $\xcom$ can be increased as (\ref{eq:invested_diffsigns}) and (\ref{eq:invested_samesigns}) show. Yet, increasing $\xcom$ makes the position switching slower as (\ref{eq:duration_switching}) shows, thus decreasing the control accuracy. The effect of $\xcom$ illustrated in the simulations of Sec. \ref{subsubsec:xcom}.

		\vspace{-0.75\baselineskip}
		\subsection{Interface circuit for the energy transfers}
		\label{subsec:interf_circuit}

\begin{figure}[htbp]
				\centering
				\includegraphics[width=.48\textwidth]{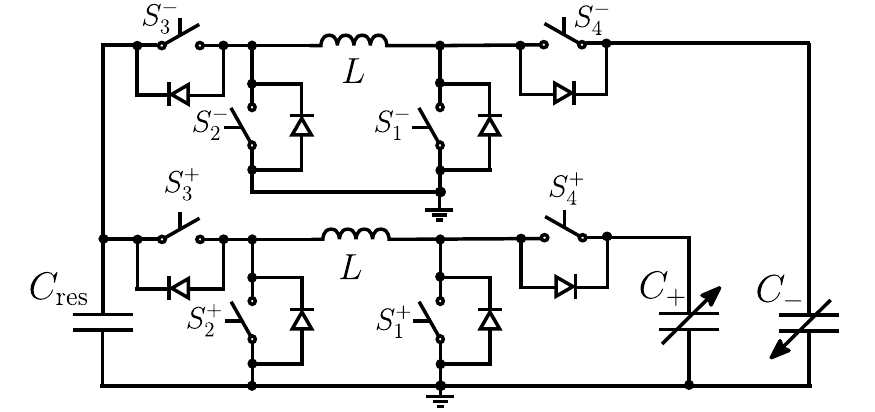}
				\caption{Interface circuit carrying out the energy exchanges to implement the control strategy described in Sec. \ref{subsec:ctrl_strategy}.}
				\label{fig:cir}
			\end{figure}

			We now propose an interface circuit to carry out the energy transfers that are needed to implement the control described in Sec. \ref{subsec:ctrl_strategy}, while allowing a recovery of the converted energy in an electrical energy tank. This circuit has a bidirectional DC-DC converter topology. Its schematic is depicted in Fig. \ref{fig:cir}. Let us describe its operation.
 			
 			In the rest of this subsection, we suppose that the transducers have fixed capacitance value during the energy transfer processes described hereafter. This implies that the processes of updating the value of the force across the transducers are fast compared to the timescale of the position switching (\ref{eq:duration_switching}), so that the mass position can be considered equal to a constant $x_0$. This hypothesis will be assessed at the end of the section.

			 \begin{figure}[htbp]
			 	\centering
			 	\includegraphics[width=.49\textwidth]{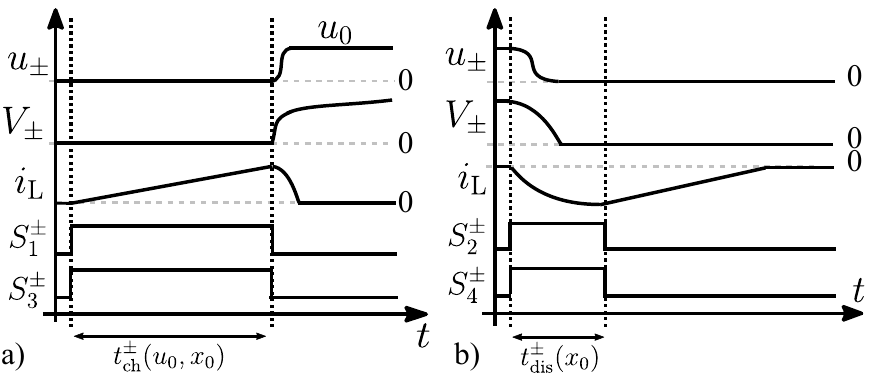}
			 	\caption{Waveforms for the interface circuit switches controls and associated electrical variables for: (a) application of a force $u_0$ across the transducer (b) nullifying the force across the transducer.}
			 	\label{fig:chronos}
			 \end{figure}

			Setting a force of value $\fz$ requires to transfer an energy $W_{u_\mathrm{0}}$ (see (\ref{eq:invested})) to the corresponding transducer ($C_+$ if \mbox{$\fz > 0$}, $C_-$ if \mbox{$\fz < 0$}). The transducer has then to be disconnected from the interface circuit, so as to operate at constant charge.
			The first step is to close switches $S_1^\pm$ and $S_3^\pm$ to transfer energy from $\Cres$ to $L$. The switches have to be kept closed for a time $t_{\mathrm{ch}}^\pm(u_0, x_0)$ to charge the inductor with the energy in (\ref{eq:invested}). If we suppose that $\Cres$ is large so that the voltage across it $\Vres$ can be considered constant, $t_{\mathrm{ch}}^\pm(u_0, x_0)$ reads:
			\begin{equation}
				t_{\mathrm{ch}}^\pm(u_0, x_0) = \frac{\sqrt{2 W_{\mathrm{u_0}} L}}{V_\mathrm{res}} = \frac{\sqrt{2 L |u_0| d_0  (1\mp x_0)}}{V_\mathrm{res}}
				\label{eq:time_chargingL_investment}
			\end{equation}

			The discharge of the transducer sets the control force to zero, while transferring the transducer energy to $\Cres$. It can be in the event of an extremum of the input $\xi(t)$, in which case the invested energy ((\ref{eq:invested_diffsigns}) or (\ref{eq:invested_samesigns})) added to the converted energy (\ref{eq:recovered}) have to be transferred from $\Cpm$ to $\Cres$. It can also be the keeping force that is being removed prior to the position switching, in which case the energy that was invested to set the keeping force has to be recovered.
			To this end, the switches $S_2^\pm$ and $S_4^\pm$ have to be closed for a time $t_{\mathrm{dis}}(x_0)$ slightly above a quarter period of the \mbox{$L \Cpm(\xz)$} cell:
			\begin{equation}
				t_{\mathrm{dis}}^\pm(x_0) = \frac{\pi}{2} \cdot \sqrt{\frac{\kappa d_0 L}{1 \mp x_0}}
			\end{equation}
			After that, the switches are opened: $L$ discharges through the diodes anti-parallel with $S_1^\pm$ and $S_3^\pm$, and then the transducer is kept at constant charge (if the leakage are negligible for the amount of time specified), generating a constant force $\fz$.

			 The waveforms depicted in Fig. \ref{fig:chronos}.a summarize the evolution of the involved switches commands and of the electrical variables during the process of updating the force from $0$ to $\fz$.
			 The waveforms depicted in Fig. \ref{fig:chronos}.b summarize the evolution of the involved switches commands and of the electrical variables during the process of updating the force from $\fz$ to $0$.

			The validity of the hypothesis that the position does not change during the force update process is now assessed. The longest force update during the mass motion is when, at $\pm\xcom$, the transducer force is updated from $0$ to $\fz$. In this case, the total duration of the force update is \mbox{$t_{\mathrm{on}}^\pm(\alpha \xi,\pm\xcom) + \pi/2\cdot\sqrt{L \Cpm(\pm\xcom)}$}. As $t^\mathrm{f}_\xi$ decreases with $\xi$ and $t_{\mathrm{ch}}^\pm(\alpha \xi,\pm\xcom)$ increases with $\xi$, the hypothesis that should be assessed is 
			\begin{equation}
				t_{\mathrm{ch}}^\pm(\alpha \xi_\mathrm{max},\pm\xcom) + \frac{\pi}{2}\cdot\sqrt{L \Cpm(\pm\xcom)} < K^\prime t^\mathrm{f}_{\xi_\mathrm{max}},
				\label{eq:hyp2}
			\end{equation}
			for large enough $K^\prime$ whose value is dictated by the tolerated inaccuracy on the constant values of the control force.




		\vspace{-0.75\baselineskip}
		\subsection{Computing, position and acceleration measurement units}
		\label{subsec:computing_sensors}

\begin{figure}[htbp]
				\centering
				\includegraphics[width=.49\textwidth]{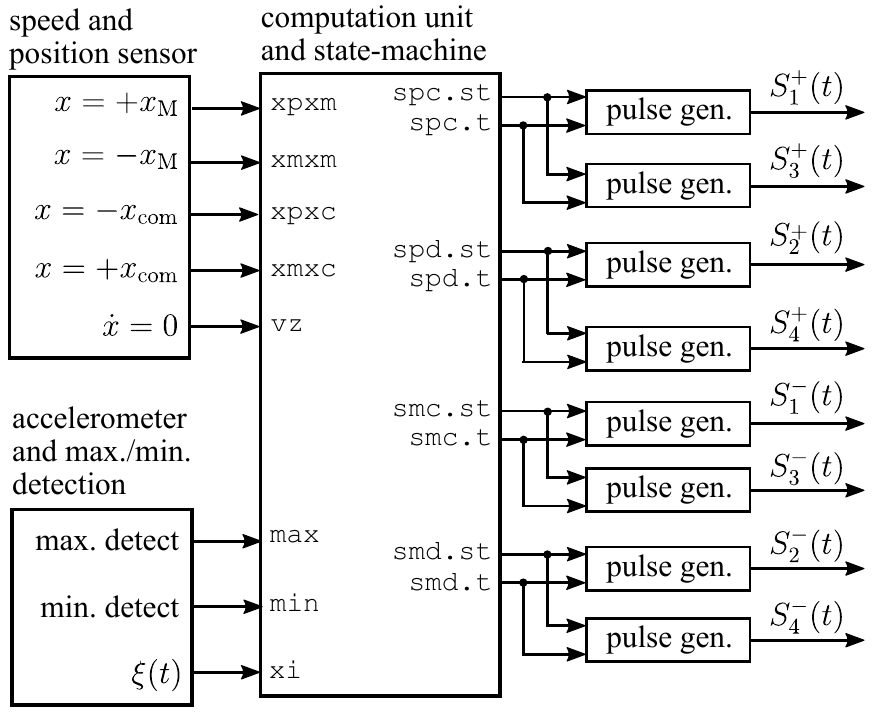}
				\caption{Architecture of the computation and sensors subsystem, that drives the interface circuit's switches to implement the control strategy of Sec. \ref{subsec:ctrl_strategy}. The regular arrows represent ``flags'', or 1-bit digital signals, whereas the crossed arrows represent quantities (that can be analog or n-bit digital depending on their exact implementation).
				}
				\label{fig:archi_ctrl}
			\end{figure}

			This section describes the sensing and computation parts of the architecture. The constitute blocks of this subsystem and their connections are depicted in Fig. \ref{fig:archi_ctrl}. To save space, and given that these blocks are present in a wide range of applications, their detailed implementation is not discussed. For simplicity, one can assume that all of the inputs and outputs in Fig. \ref{fig:archi_ctrl} and that are mentioned below are digital.

%

			Note that the control $u(t)$ as described in Sec. \ref{subsec:ctrl_strategy} does not require a measurement of the velocity. However,  the "computation and state-machine" of Fig. \ref{fig:archi_ctrl} needs a zero velocity detector (the $\dot x=0$ input. This is done in order to increase the robustness of the control. Indeed, because of control inaccuracy, the mass velocity may vanish before it reaches the targeted end-stop in the position commutation and return to the end-stop that it has left when the extremum was detected. This would result in losing track of the mass position. To overcome this, when the velocity vanishes during the position commutation, the keeping force is applied so as to move the mass to the targeted end-stop (inducing end-stop collision losses and/or deviation from the optimal trajectory).

\begin{figure}[htbp]
				\centering
				\includegraphics[width=.49\textwidth]{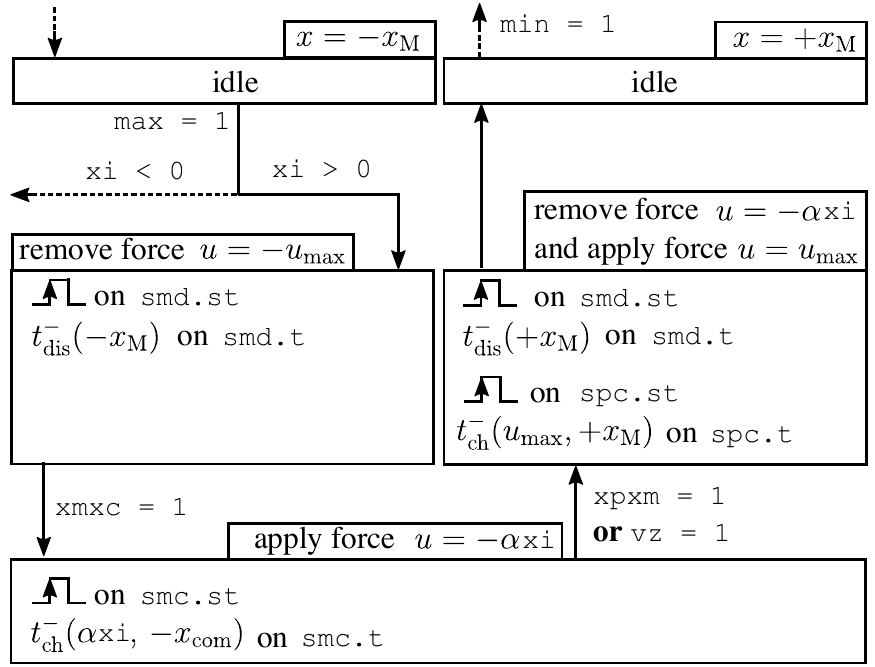}
				\caption{Schematic of the finite state automaton. The part of the automaton that is depicted details the sequence for the position switching from $-\xm$ to $+\xm$ when a maximum of positive value is detected. The dotted arrow are connected to states that are not involved in this position switching event.
				}
				\label{fig:mae}
			\end{figure}

\begin{figure}[htbp]
				\centering
				\includegraphics[width=.49\textwidth]{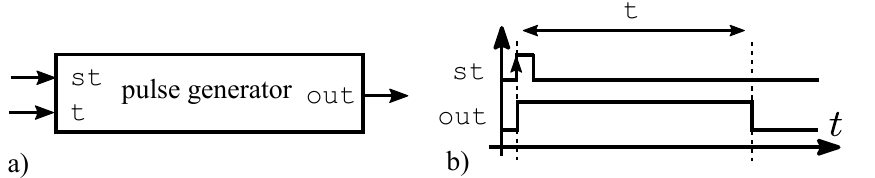}
				\caption{(a) Pulse generator block. (b) Waveform describing its operation.}
				\label{fig:pulse_generator_block}
			\end{figure}

			The value of the input acceleration and the flags corresponding to the position and velocity of interest, and to the maximum/minimum detection, are fed into the main computation unit. This unit computes interface circuit's switches activation signals needed to implement the control force. To this end, it incorporates a finite-state automaton that sets the value of the output according to the sensed events. The graph of this finite-state automaton is depicted in Fig. \ref{fig:mae} for the case of a sensed maximum with \mbox{$\xin x(t_\mathrm{n}) < 0$}. 
			The outputs of the finite-state automaton are a time value and a pulsed command for each switch of the electrical interfacing circuit. These outputs are fed into a programmable pulse generator. This component generates waveforms corresponding to the delays specified on its input \texttt{t}, when it is activated by a rising edge signal on its input \texttt{st}.  By default, when the component is not generating any waveform, the output is zero.  The detail of the pulse generation operation is depicted in Fig. \ref{fig:pulse_generator_block}.

			A circuit that is not depicted in Fig. \ref{fig:archi_ctrl} is needed in order to synthesize the suited driving signals for the switches of the interfacing circuit, from the outputs of the pulse generators. This circuit depends on the chosen technology for the switches (e.g., gate level shifter for power MOSFET switches).


	\vspace{-0.5\baselineskip}
	\section{Sizing and simulations of the near-limits electrostatic KEH architecture}
	\label{sec:sizing_simulations}

		In this section, the architecture described in Sec. \ref{sec:electrostatic_implementation} is tested in simulation. The simulations are carried out using Cadence AMS, which allows coupled simulation of the mechanical part described in terms of ODEs, the electrical part using a SPICE engine solver, and the computation/sensing part using behavioral VHDL-AMS models \cite{galaykovhdlams2007}. The modeling and parameters choice are the subject of Sec. \ref{subsec:sizing}. The results of the simulations for the sized architecture are given in Sec. \ref{subsec:simulations_illustration}. Then, the impact of some of the architecture's parameters on the harvested energy is highlighted in Sec. \ref{subsec:levers_optimization}.

		The mechanical input excitation used for the simulations is a $\SI{1}{\second}$ sample extracted from a longer recording of the acceleration on a running human using a smartphone accelerometer. The sample is representative of the recorded signal in its full duration, and corresponds roughly to two steps of running. The acceleration is recorded following the person's height axis. The smartphone was put in the trousers front pocket.

		\vspace{-0.75\baselineskip}
		\subsection{Modeling and sizing of the architecture}
		\label{subsec:sizing}

			In this section, the architecture is parametrized for the simulation. Let us put the sole loose constraint that the total volume of the KEH should be of the order of \mbox{$\SI{1}{\centi\meter\cubed}$}, and that the mechanical device is made of silicon. The mechanical resonator and the associated differential electrostatic transducer are modeled in VHDL-AMS, similarly to what is reported in \cite{galaykovhdlams2007}. The interface circuit is described and simulated as a SPICE netlist for the coupled simulation. 

			\subsubsection{Mechanical part}
			\label{subsubsec:sizing_mechanical}

				For the imagined gap-closing electrostatic transducer, a geometry similar to the structure depicted in Fig. \ref{fig:electrostatic_nearlimits_keh} is chosen. It is a cuboid-shaped mass sliding in between two electrodes, to which it is connected by spring suspensions. We select a mass of \mbox{$m = \SI{1}{\gram}$}, and a spring stiffness of \mbox{$k = \SI{400}{\newton\per\metre}$}, yielding a resonant frequency of \mbox{$\SI{100}{\hertz}$} for the subsequent resonator. The transducer surface is of \mbox{$S = \SI{10}{\centi\meter\squared}$}, and the gap at \mbox{$x = 0$} is of \mbox{$d_0 = \SI{50}{\micro\meter}$}. A device with such parameters can be obtained, e.g., by a differential and slightly upscaled version of the transducer structure reported in \cite{mitchesonharvesting2008} fabricated in silicon. With this geometry, the active area is of \mbox{$\SI{3.16}{\centi\meter}\times\SI{3.16}{\centi\meter}$} for a depth of \mbox{$\SI{450}{\micro\meter}$}. The mechanical device's volume is then of \mbox{$\SI{0.55}{\centi\meter\cubed}$}. The stoppers are placed so that $X_M=0.95$ (see Fig. \ref{fig:electrostatic_nearlimits_keh}). The maximum and minimum values of the transducer's capacitances are respectively \mbox{$\Cpm(\pm\xm) = \SI{3.54}{\nano\farad}$} and \mbox{$\Cpm(\mp\xm) = \SI{90}{\pico\farad}$}. The mass, seen as a beam of width \mbox{$\SI{450}{\micro\meter}$}, length \mbox{$\SI{3.16}{\centi\meter}$} and height \mbox{$\SI{3.16}{\centi\meter}$}, has a stiffness of \mbox{$\SI{17.5}{\kilo\newton\per\metre}$} in the direction normal to the transducer's electrode plane (considering a Young modulus of \mbox{$\SI{192}{\giga\pascal}$} for silicon). Given the value of $\xi_\mathrm{max}$, this value allows  considering that the mass deflection in the direction normal to the transducer's electrode plane is negligible, so that the lumped mass model used to build the harvester's architecture is accurate. A linear damping effect is incorporated, to model an air friction phenomenon. The corresponding damping coefficient is \mbox{$\mu = \SI{0.2}{\milli\newton\per\meter\second}$}, corresponding to a quality factor of \mbox{$Q = 10$} for the obtained linear resonator. Note that non-linear air damping effects can have significant impact on the dynamics of small-sized KEHs. Yet, vacuum-packaging \cite{elfrink2010vacuum} or clever transducer geometry design \cite{lu2017new} have shown to greatly reduce its effect.
				The mechanical end-stops were modeled inspired by \cite{truong2015verified}: a linear spring \mbox{$k_\mathrm{st} = \SI{1}{\mega\newton\per\metre}$} associated with a linear damper \mbox{$\mu_{\mathrm{st}} = \SI{10}{\newton\second\per\metre}$}. 

				For the first simulations, $\xcom$ is selected as \mbox{$\xcom = \xm/2$} ($\xcom$ is then varied in Sec. \ref{subsubsec:xcom}). Therefore, \mbox{$\Cpm(\pm\xcom) = \SI{337}{\pico\farad}$}. In order estimate the minimum position switching time as well as to chose the keeping force $u_\mathrm{max}$, an estimate for $\xi_{\mathrm{max}}$ is needed. Let us suppose that \mbox{$\xi_\mathrm{max} \leq \SI{60}{\milli\newton}$}. Therefore, $t_\xi^\mathrm{f}$ ranges from \mbox{$\SI{2.84}{\milli\second}$} to \mbox{$\SI{4.97}{\milli\second}$}. The keeping force must be chosen greater than \mbox{$\xi_\mathrm{max} + k d_0 \xm$}. We choose \mbox{$u_\mathrm{max} = \SI{75}{\milli\newton}$}.


			\subsubsection{Electrical part}
			\label{subsubsec:sizing_electrical}

		 		The electrical energy tank is modeled as a large ideal capacitor \mbox{$\Cres = \SI{10}{\micro\farad}$} and with \mbox{$\Vres = \SI{10}{\volt}$}.

				The switches $S_1^\pm$ and $S_2^\pm$ block the transducer voltage, that can go to up to \mbox{$\SI{415}{\volt}$} as indicated by (\ref{eq:vmax}), for \mbox{$\xcom = \xm/2$}. These four switches are modeled as power MOSFETs with their anti-parallel body diodes. They have a major impact on both the control generation cost and accuracy. We choose to parametrize them such that their parasitic capacitance at zero blocking voltage \mbox{$C_{\mathrm{J 0}} = \SI{10}{\pico\farad}$}. This value is selected as the largest $C_{\mathrm{J 0}}$ in the range \mbox{$\SI{5}{\pico\farad}$} to \mbox{$\SI{50}{\pico\farad}$} (varied by steps of \mbox{$\SI{2.5}{\pico\farad}$}) that yields less than $5\%$ error on the transducer force when all resistive losses are compensated, for forces values up to \mbox{$\SI{100}{\milli\newton}$} and for a transducer capacitance of $\Cpm(\pm\xcom)$. From the relations provided in \cite{mitcheson2012maximum} for power MOSFET blocking up to \mbox{$\SI{420}{\volt}$}, these switches are parametrized such that their ON-state resistance is of \mbox{$R_{\mathrm{ON}} \leq \SI{19}{\ohm}$}.

		 		To fulfill the size requirement of the KEH, inductors of size that do not exceed \mbox{$\SI{0.25}{\centi\meter\cubed}$} are considered. Given the selected value for $R_\mathrm{ON}$, \mbox{$L = \SI{1}{\milli\henry}$} (modeled as an ideal inductor element in series with its parasitic DC resistance of \mbox{$R_\mathrm{L} = \SI{6}{\ohm}$} \cite{wurthinductor}) is nearly optimal regarding to the resistive energy losses. The electrical timescale of the force update can be estimated: \mbox{$t_{\mathrm{ch}}(\alpha\xi_\mathrm{max},\xcom) + \pi/2\cdot\sqrt{L C_\pm(\pm \xcom)} = \SI{6.5}{\micro\second}$}. 

		 		The parts that were described in Sec. \ref{subsec:computing_sensors} are modeled in VHDL-AMS as behavioral blocks, as are the gate level shifters driving circuit's MOSFETs.

		\vspace{-0.75\baselineskip}
 		\subsection{Simulation illustrating the architecture's operation}
 		\label{subsec:simulations_illustration}

			\begin{figure}[htbp]
				\centering
				\includegraphics[width=.49\textwidth]{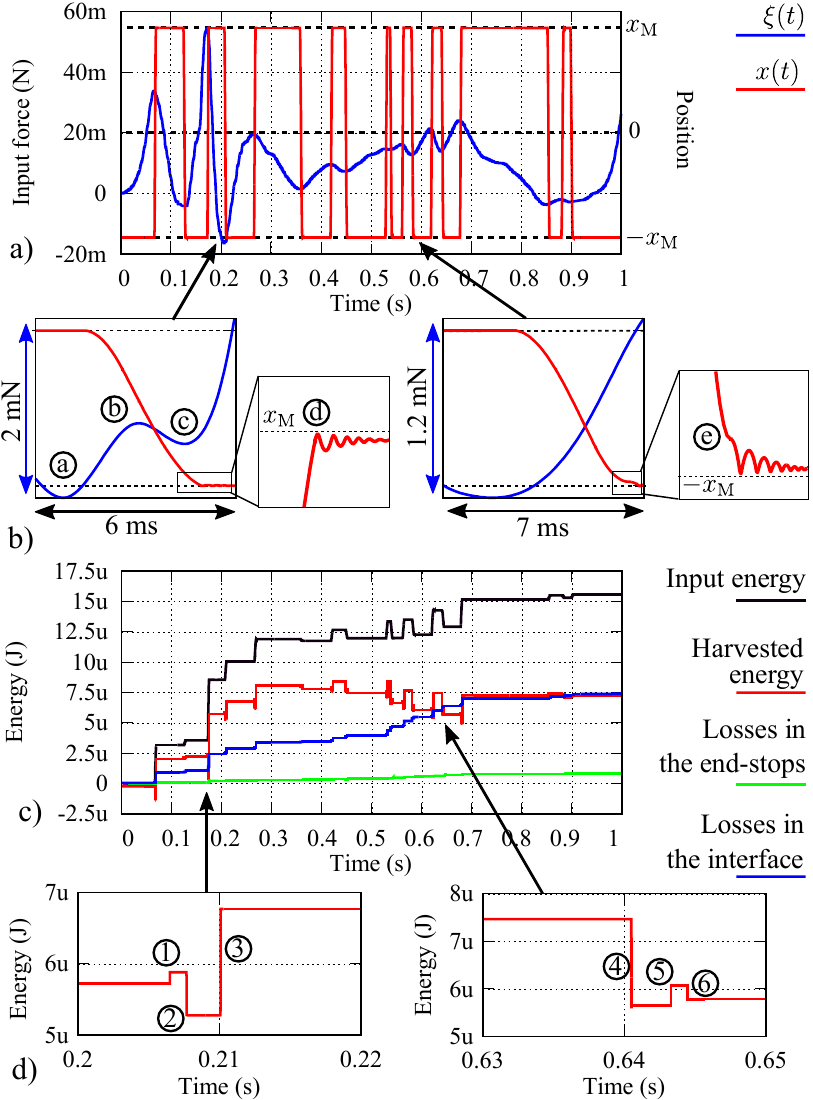}
				\caption{Results of the simulation of the near-limits electrostatic KEH --- (a) Mass position and input force. --- (b) Magnification around two extrema: the minimum at \textcircled{a} is detected and hence the control makes the mass move from $+\xm$ to $-\xm$. The extrema at \textcircled{b} and \textcircled{c} are ignored as they happen during the position switching. At \textcircled{d}, the control inaccuracy provokes a collision with the end-stop at $+\xm$, since the mass reaches this position with non-zero velocity. At \textcircled{e}, the control inaccuracy makes the mass velocity vanish before $-\xm$, so the control unit applies $-u_\mathrm{max}$ and the mass collides with the end-stop at $-\xm$. --- (c) Energies in the KEH. The harvested energy refers to the difference of the energy in $\Cres$ between times \mbox{$t=0$} and $t$. --- (d) Magnification of the harvested energy around an extremum that is such that \mbox{$\xi_0 x(t_0) < 0$}. At \textcircled{1}, energy is gathered back from removing $-u_\mathrm{max}$, and the mass starts travelling towards $+\xm$ as $\xi(t)$ acts upon it. At \textcircled{2}, energy is invested to implement the control force working negatively on the mass. At \textcircled{3}, the sum (invested energy + converted energy - energy to implement $u_\mathrm{max}$ at $+\xm$) is gathered back in $\Cres$. (e) Magnification of the harvested energy around an extremum that is such that \mbox{$\xi_1 x(t_1) > 0$}. At \textcircled{4}, the energy to implement the negatively working control force is invested, while some energy is being recovered from removing $u_\mathrm{max}$ at $+\xm$. At \textcircled{5}, part of the invested energy is recovered and the mass continues to travel towards $\xm$. At \textcircled{6}, some energy is invested from $\Cres$ to implement $u_\mathrm{max}$.}
				\label{fig:sim_results}
			\end{figure}



			Results illustrating the operation of the system submitted to the $\SI{1}{\second}$ input excitation are depicted in Fig. \ref{fig:sim_results}. The plot in Fig. \ref{fig:sim_results}.a shows the mass trajectory and the input force. A magnified plot of both, explicating the mass position switching around two extrema of $\xi(t)$, is depicted in Fig. \ref{fig:sim_results}.b. 

			The evolution of the harvested energy is depicted in Fig. \ref{fig:sim_results}.c. At the end of the sample, it is equal to \mbox{$\SI{7.2}{\micro\joule}$}, whereas the system input energy is of \mbox{$\SI{15.6}{\micro\joule}$}. The sample plot shows the evolution of the total energy lost in the elements of the interface circuit, which amounts to \mbox{$\SI{7.3}{\micro\joule}$}. The losses in the stoppers, reflecting the control imprecision, amount to \mbox{$\SI{0.8}{\micro\joule}$}. For the energy balance to be complete, one has to consider the instantaneous energy stored in $\Cpm$, in the inductors, in the spring and the energy dissipated by the linear damper accounting for air friction. As these sum up to smaller amounts than the other energies, they are not represented in the energy balance in Fig. \ref{fig:sim_results}.c.

			The results show that nearly as much energy is harvested as is lost in the electrical interface. The losses in the end-stops are relatively small, as accuracy in the force was privileged over resistive losses when selecting the electrical components in Sec. \ref{subsubsec:sizing_electrical}. The good accuracy is further reinforced by the relative values of the electrical and mechanical timescales. Let us now discuss how some slight modifications in the control law can help increase the harvested energy.

		\vspace{-0.75\baselineskip}
		\subsection{Control-related levers for optimization}
		\label{subsec:levers_optimization}


			\subsubsection{Extremum selection}
			\label{subsubsec:extremum_selection}

				The results in Fig. \ref{fig:sim_results}.b show that the consecutive extremum happening between \mbox{$\SI{0.5}{\second}$} and \mbox{$\SI{0.7}{\second}$} lead to an overall decrease in the harvested energy. This is because the cost of carrying out the position switchings overcomes the converted energy for these pairs of extrema. In such a case, it can be beneficial to skip some extrema.

				\begin{figure}[thbp]
					\centering
					\includegraphics[width=.49\textwidth]{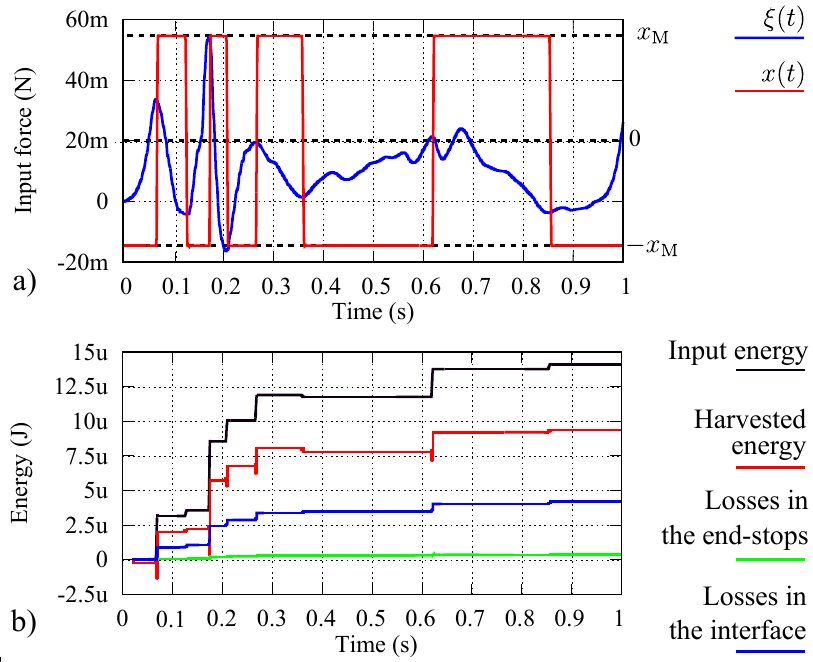}
					\caption{Results of the simulation of the near-limits electrostatic KEH, with a threshold for extremum detection of $\SI{15}{\milli\newton}$.}
					\label{fig:sim_results_tol}
				\end{figure}

				To this end, a primitive criterion of extremum selection for the position switching can be used. It consists in carrying out the position switching only if the difference with the previous extremum exceeds a threshold value, that we select equal to $\SI{15}{\milli\newton}$. This results in skipping the extrema that led to a decrease of the harvested energy at the end of the sample.

				The system is then simulated, and the results are depicted in Fig. \ref{fig:sim_results_tol}. The trajectory deviates from $x_\xi(t)$ in that the extremum happening between \mbox{$\SI{0.4}{\second}$} and \mbox{$\SI{0.8}{\second}$} are now skipped, since they cost is higher then their contribution to the harvested energy (see Fig. \ref{fig:sim_results}.b). Compared to the previous simulation without a selection threshold, this results in a decreased input energy (\mbox{$\SI{14.1}{\micro\joule}$}) but increased harvested energy (\mbox{$\SI{9.4}{\micro\joule}$}). Both the losses in the electrical interface (\mbox{$\SI{4.2}{\micro\joule}$}) and in the end-stops (\mbox{$\SI{0.3}{\micro\joule}$}) are decreased. This is an obvious example where a different trajectory than $x_\xi(t)$ yields higher harvested energy than $x_\xi(t)$.

                    Note however that the maximum at \mbox{$\SI{0.62}{\second}$} is selected, whereas it would have been more beneficial to skip it and select the next maximum at \mbox{$\SI{0.67}{\second}$}. Nonetheless, increasing the threshold value so as to select the latter would in fact make the extremum at \mbox{$\SI{0.35}{\second}$} to be skipped. The subsequent sequence of selected extremum would then result in a decreased amount of harvested energy. 
                    
                    To further maximise the net energy, one can use smarter algorithm able to predict the next (future) extrema of the $-ma_{ext}$ function. The prediction can be made thanks to existence of internal correlations present in the vibrations. Indeed, vibrations in real systems usually follow some repeating patterns (contrarily to white model noise models widely used in theoretical studies \cite{scruggs2012multi,scruggs2010causal,halvorsen2008energy}), and an efficient prediction  may be achieved by standard tools of time series analysis.

			\subsubsection{$\xcom$ parameter}
			\label{subsubsec:xcom}

				\begin{figure}[htbp]
					\centering
					\includegraphics[width=.49\textwidth]{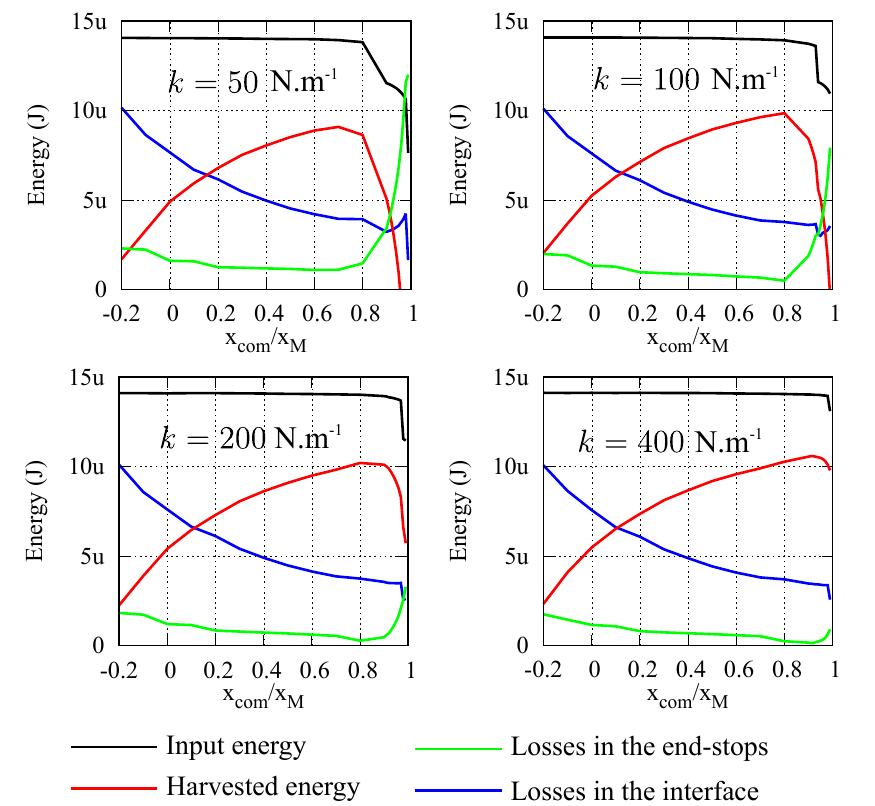}
					\caption{Simulation results for the near-limits KEH, varying $\xcom$ and $k$. Each point corresponds to the energies at the end of the sample.}
					\label{fig:sim_results_xcom}
				\end{figure}

				The role of the control parameter $\xcom$ was discussed in Sec. \ref{subsubsec:energy_considerations_ctrl}. In addition to the decreased invested energy, increasing $\xcom$ allows choosing switches with reduced resistance and capacitance, thanks to the reduced maximum transducer voltage (see (\ref{eq:vmax})). Another advantage in increasing $\xcom$ is that, for extrema such that \mbox{$\xi_n x(t_n) < 0$}, the non-zero force is applied when the transducer's capacitance is $C_\pm(\pm\xcom)$ which increases with $\xcom$. Thus, the inaccuracy on the force value is decreased, as the transducer capacitance is larger compared to the switches capacitances.

				The simulation results depicted in Fig. \ref{fig:sim_results_xcom} show the energy balance at the end of the $\SI{1}{\second}$ input, for values of $\xcom$ ranging from $-\xm$ to $\xm$ (excluded), and for different $k$. The  selection threshold of \mbox{$\SI{15}{\milli\newton}$} is used. For each point, new values of $R_\mathrm{ON}$ and $C_\mathrm{JO}$ are used, which yield the highest harvested energy for the corresponding maximum transducer voltage.

				The results are slightly improved compared to Sec. \ref{subsubsec:extremum_selection} (\mbox{$k = \SI{400}{\newton\per\meter}$}). The harvested energy is increased from \mbox{$\SI{9.4}{\micro\joule}$} with \mbox{$\xcom = 0.5\xm$} to \mbox{$\SI{10.6}{\micro\joule}$} with \mbox{$\xcom = 0.9\xm$}. This is the highest value of harvested energy obtained from this sample. It amounts to $68\%$ of the absolute limit in \emph{input energy} that can be obtained with the considered mechanical device and input excitation. This value of $\xcom$ yields a maximum transducer voltage of \mbox{$\SI{370}{\volt}$}, so that $R_\mathrm{ON}$ is reduced from $\SI{19}{\ohm}$ with \mbox{$\xcom\!=\!\xm/2$}, to \mbox{$R_\mathrm{ON} = \SI{15}{\ohm}$} with \mbox{$\xcom\!=\!\xm/2$} \cite{mitcheson2012maximum}.

				When $k$ is large, increasing $\xcom$ is beneficial up to values that are close to $\xm$. This is because large $k$ ensures that the position switching remains fast enough, despite the increase of $\xcom$, so that the control effectively implements the ideal trajectory $x_\xi(t)$. As $k$ is decreased, the position switching becomes slower. Hence, the value of $\xcom$ for which the control becomes inaccurate and the trajectory deviates from $x_\xi(t)$ is reduced. The decrease in the input energy for values of $\xcom$ approaching $\xm$ is due to a significant deviation of $x(t)$ from $x_\xi(t)$. For example, with \mbox{$k=\SI{50}{\newton\per\meter}$} and \mbox{$\xcom = 0.9\xm$}, the switching time for the maximum at \mbox{$\SI{0.17}{\second}$} (the largest extremum of the example) is of \mbox{$\SI{6.8}{\milli\second}$}, during which $\xi(t)$ decreases by $30\%$ from its value at the extremum.

		\vspace{-0.75\baselineskip}
		\subsection{Discussion}
			
			The sizing of the near-limits KEH done above started by considering a $\SI{1}{\centi\meter\cubed}$ size constraint was considered. A mechanical device inspired by existing designs and compatible with this constraint was chosen \emph{a priori} (Sec. \ref{subsubsec:sizing_mechanical}). Only then, the electrical part was optimized under a transducer force accuracy constraint (Sec. \ref{subsubsec:sizing_electrical}), hence predefining the balance between the control accuracy and the energy losses in the interface circuit. This determined value of the resistive components of the circuit elements, which are the origin of a major component of energy loss in the electrical interface. Finally, after both the mechanical part and the electrical part were fixed, the control was tuned in order to maximize the harvested energy (Sec. \ref{subsec:levers_optimization}).

			In fact, allowing for customized sizing of the mechanical device and for different balances between the control accuracy and the losses in the electrical interface may improve the figures of harvested energy, instead of the sequential design choices described above. The subsequent optimization should include parameters related to the control, to the mechanical part, and to the circuit components, under constraints that are dictated by the application (e.g., maximum size constraint) and for known characteristics of the input (at least $\xi_\mathrm{max}$). This can be done by enriching the problem $(Q)$ with cost functionals related to the chosen mechanical and electrical parts.

			Optimization levers in electrical interface include, e.g., the value of $\Vres$. It can be shown that increasing $\Vres$ decreases the dissipated energy in the interface circuit's elements, but can may necessitate a step-down conversion mechanism if the voltage required by the supplied electronics is small. Also, note that other technologies than power {MOSFET} may offer reduced dissipation \cite{stark2005comparison}. Detailed electrical interface optimization for kinetic energy harvesters can be found in, e.g., \cite{mitcheson2012maximum,tabesh2010low}, which deal with electrical interfaces similar to that of our architecture, albeit without the additional optimization levers and requirements posed by the trajectory control of the near-limits KEH architecture.

			Finally, note that a limitation of the above simulations is that they do not take into account the consumption of the blocks of Sec. \ref{subsec:computing_sensors}, as they were described using behavioral models. Some of these parts have to be taken into account in the overall system optimization, as their energy consumption depends on the system's parameters. For example, the temporal resolution of the position detector should be low enough compared to the position switching time, which goes with increased power consumption. This may mitigate the results of Fig. \ref{fig:sim_results_xcom} which show the benefits of increasing $k$, because larger $k$ decreases the position switching time (see (\ref{eq:duration_switching})) and hence puts more stringent requirements on the position detector's temporal resolution. Likewise, the pulse generator and computation speeds have be negligible with respect to the position switching time. State-of-the-art realizations of each of the blocks constituting the computing and sensing part show micro-watt or sub-micro-watt energy consumption figures \cite{stLIS2DW12,tiads7042,gasnier2013maxdetect,jazairli2010ultra,jung201527}. This is comparable to the harvested energy, further reinforcing the need for a fine system optimization.

	\vspace{-0.5\baselineskip}
	\section{Conclusion}
	\label{sec:conclusion}

		The architecture of an electrostatic KEH intended to maximize the harvested energy from arbitrary vibration sources was described in detail. Simulation results support the proof of concept of such a system and show the effect of some of the architecture's parameters that can be used to optimize the KEH's operation. A figure of harvested energy that amounts to up to $\SI{10.6}{\micro\joule}$ is attained, from a $\SI{1}{\second}$ vibration sample recorded on a running human. This amounts to $68\%$ of the absolute limit set by the device size and the input. Although the consumption of the parts responsible for the computation and sensing was not taken into account, the architecture still has room for improvement by the mean of a full-system optimization. Future works may focus on such an optimization, as well as on a practical implementation of a near-limits KEH based on the presented architecture.
		%



	\vspace{-0.5\baselineskip}

	\ifCLASSOPTIONcaptionsoff
		\newpage
	\fi

	\bibliographystyle{IEEEtran}
  	\bibliography{draft.bib}

	\IEEEpeerreviewmaketitle


\end{document}